# Evidence for the intrinsic ferromagnetism of semiconductor (Ga,Cr)As revealed by magnetic circular dichroism

H. D. Gan,[a)] H. Z. Zheng,[b)] J. F. Bi, Y. Ji, B. Q. Sun, G. R. Li, X. Z. Ruan, J. Lu, J. H. Zhao

*State Key Laboratory for Superlattices and Microstructures, Institute of Semiconductors, Chinese Academy of Sciences, P. O. Box 912, Beijing 100083, China*

Abstract

To clarify whether or not (Ga,Cr)As is an intrinsic diluted magnetic semiconductor, a systematic study of the magnetic circular dichroism (MCD) was carried out for a series of (Ga,Cr)As epilayers grown by the low-temperature molecular-beam epitaxy technique. The present work provides unambiguous evidence for the intrinsic ferromagnetism of the (Ga,Cr)As epilayers with all the necessary properties in accordance with that of a diluted magnetic semiconductor, especially the hysteresis characteristics, which is opened up in the magnetic field dependence of MCD.

[a)]Electronic mail: hdgan@red.semi.ac.cn

[b)]Electronic mail: hzzheng@red.semi.ac.cn

The most important issue in diluted magnetic semiconductors (DMSs) research is to incorporate the function of localized $d$ electrons with that of $sp$ band electrons in a single semiconductor. As a result, new devices for processing and storing information on an equal footing could be invented [1]. Kinetic mixing between $sp$ band electrons and localized $d$ electrons not only plays a dominant role in modifying the band structure of host semiconductors by giant Zeeman splitting, but is also a key mechanism responsible for the hole-mediated ferromagnetic coupling between magnetic ions in DMSs. Since the magnetic circular dichroism (MCD), produced either by the presence of the magnetization in materials or induced by spin split band structures, is an effect in which the reflection and transmission coefficients of DMSs are different when probed by clockwise-polarized ($\sigma^+$) and counterclockwise-polarized ($\sigma^-$) lights, the dual roles of the $sp$-$d$ exchange interaction are most reliably confirmed by MCD for diluted magnetic semiconductors [2].

In contrast to the MCD signal of a conventional magnetic material, the MCD spectra of ferromagnetic semiconductors show a close correlation with the band structure of the parent semiconductor. In particular, when the magnetic field dependence of its MCD signal displays the behavior of ferromagnetic hysteresis, the intrinsic ferromagnetism of a MCD can be convincingly verified. Until now, MCD analysis has only verified that (In,Mn)As, (Ga,Mn)As, and (Zn,Cr)Te are intrinsic diluted magnetic semiconductors [3-5]. Debate still remains whether or not GaN:Mn and ZnO:Ni are true intrinsic diluted magnetic semiconductors, because either the MCD signal has been not observed or is not compatible with what is expected from the sp-d exchange interaction [6,7]. Regarding (Ga,Cr)As, theoretical calculations predicted a stronger ferromagnetism and higher $T_C$ than (Ga,Mn)As [8]. However, experimental results revealed only inconsistencies with theoretical predictions. (Ga,Cr)As initially displayed superparamagnetism [9]. Then, it was thought that the short-range ferromagnetic coupling was probably suitable for describing the magnetism of (Ga,Cr)As on the basis of the distributed magnetic-moment model [10]. Recently, ferromagnetic (Ga,Cr)As films were reported in the presence of the second precipitates [11]. Accordingly, whether or not (Ga,Cr)As is an intrinsic diluted magnetic semiconductor is still an open question. To clarify these contradictions, we carried out a detailed MCD study for (Ga,Cr)As epilayers grown by low temperature molecular beam epitaxy (LT-MBE), and provided unambiguous evidence for its intrinsic ferromagnetism as a diluted magnetic semiconductor.

A series of (Ga,Cr)As samples in investigation were grown by LT-MBE at 250 $^{\circ}$C with different Cr concentrations and V/III flux ratios. The layer structure was composed of a 100-nm-thick (Ga,Cr)As epilayer grown on semi-insulating GaAs (001)substrate. A clear ($1 \times 1$) reflection high energy electron diffraction (RHEED) pattern was observed when Cr% < 5%, but the RHEED pattern became spotty for Cr% > 5%. The crystalline quality was routinely examined by double crystal x-ray diffraction measurements. Fig. 1 (a) shows that the (Ga,Cr)As peak is clearly separated from the GaAs peak in the samples of 1% and 1.5% Cr under the (004) Bragg reflection with lattice mismatches of 0.14% and 0.07%, respectively. When the Cr mole fraction was increased to 3% and 5%, the scattering-like contribution from the (Ga,Cr)As layer appears as a shoulder on the left side of the GaAs (004) peak indicating poorer crystalline quality.

Fig.1 (b) gives the temperature dependence of the in-plane remanent magnetization ($M_r$) of the (Ga,Cr)As sample with Cr% = 1%, measured by a superconductor quantum interference device (SQUID). A ferromagnetic remanent magnetization is observed at temperatures below ~12 K. As the temperature rises above it, the remanent magnetization rapidly decreases to zero and no residual magnetization appears from possible second phase precipitates. The conspicuous hysteresis loop also shows up in the magnetic field dependence of the magnetization, as shown in the inset of Fig. 1(b), indicating the ferromagnetic behavior in the (Ga,Cr)As sample with 1% Cr. These facts suggest the single-phase ferromagnetism of the (Ga,Cr)As epitaxial film. However, that the observed ferromagnetism is an intrinsic behavior of the diluted magnetic semiconductor needs to be verified by a method other than SQUID magnetization characteristics. For this purpose, the MCD spectrum was measured in the reflection configuration within a spectrum range from 500 to 1000 nm. The monochromatic light source was provided by passing a supercontinuum-generation white light [12] through a spectrometer. Alternating circular polarization lights (50 kHz) are modulated by passing the light through a fused-silica photo-elastic modulator (PEM). Then, the polarization-modulated light was focused almost perpendicularly on the sample and in parallel to the applied magnetic field. The focused spot on the sample had a diameter of about 1 mm. The reflected light deviated from the incident light by 5° and was detected by a Si-photodiode detector. The MCD signal was measured by a Lock-in amplifier.

Since the intensities of the $\sigma^+$ and $\sigma^-$ reflected lights are measured in the reflection configuration, the MCD signal is proportional to the imaginary part of the non-diagonal matrix element of the dielectric susceptibility $\chi_{\mu\nu}(\omega)$, as given by

$$MCD = \frac{90}{\pi}\frac{R^+ - R^-}{R^+ + R^-} = -\frac{180}{\pi}\frac{\chi_{xy}^{''}}{n(n^2-1)} \qquad (1).$$

From the Kubo formula for the dielectric susceptibility [13], it is given as

$$\chi_{xy}^{''}(\omega) = -\frac{1}{4V\hbar}\sum_{v,c}(\rho_v - \rho_c) \times \left(\left|M_+\right|_{vc}^2 - \left|M_-\right|_{vc}^2\right)\frac{(\omega_c - \omega_v - \omega)}{(\omega_c - \omega_v - \omega)^2 + \Gamma^2} \qquad (2),$$

where $\rho_c$ and $\rho_v$ are the density functions of the conduction and valence bands, respectively, and $\Gamma$ is the effective broadening parameter. $\left|M_+\right|_{vc}^2$ and $\left|M_-\right|_{vc}^2$ are the squared dipolar transition matrix elements for $\sigma^+$ and $\sigma^-$ polarizations. In the framework of Kane $k$-$p$ perturbation theory, $\left|M_+\right|_{vc}^2$ and $\left|M_-\right|_{vc}^2$ near the critical point ($\Gamma$) increase approximately in proportion to $E_g^+$ and $E_g^-$, respectively. Here $E_g^+$ and $E_g^-$ are the interband transition gaps for two optical polarizations, $\sigma^+$ and $\sigma^-$. Because the giant Zeeman splitting induced by antiferromagnetic p-d exchange interaction usually makes $E_g^+$ smaller than $E_g^-$, the MCD signal has a negative sign when the photon energy is above the band gap. The Lorentz form of the third term in Eq. (2) forces the MCD signal to decrease rapidly as $\omega$ deviates from the band gap.

Following the above formulism, the MCD spectra of the (Ga,Cr)As sample with $x$ = 1% have been measured at $T$ = 1.5 K and $B$ = 1 T in Fig. 2 in order to check their correlation with the band structure. The lower-left inset of Fig. 2 displays a pronounced MCD signal in a spectrum range from 600 nm to 960 nm. The polarity of the measured MCD signal was checked with respect to the magnetic field dependence of the static intensity difference between $\sigma^-$ and $\sigma^+$ polarization light measured in the absence of PEM. The measured polarity of MCD is negative consistent with what is expected from the above expression. The most noticeable feature in Fig. 2 is the hysteresis in the magnetic field dependence of MCD and its rapid saturation. The hysteresis shown in MCD vs. $B$ unambiguously proves that the observed ferromagnetic signal comes from the intrinsic nature of (Ga,Cr)As, when it acts as DMS. Moreover, the different hysteresis loops, measured at different wavelength $\lambda = 800, 820, 860$ and $890$ nm, collapse onto the same curve, after being normalized by the MCD value at $B = 1$ T. In Fig. 2, only two of them ($\lambda = 800, 890$ nm) are plotted for the sake of clarity. The independence of MCD on the wavelengths near the $\Gamma$ critical point is also well explained by the theory for interband dielectric susceptibility in a DMS. The saturated MCD in its magnetic field dependence actually reflects the magnetization characteristics. The upper-right inset of Fig. 2 compares the magnetic field dependence of the MCD intensity ($\lambda = 860$ nm) with that of the magnetization measured by SQUID. Their excellent coincidence again implies that the magnetization of our (Ga,Cr)As epilayer stems from the hole-mediated ferromagnetic coupling between magnetic ions, where $p$-$d$ exchange interaction plays a central role. No second-phase ferromagnetic precipitate exists. We made an additional check by sweeping the magnetic field from 1 T to 0 T and from -1 T to 0 T. For both cases, MCD spectral features at zero field reverse their signs, indicating the existence of the giant Zeeman splitting even at zero field.

To further ascertain the ferromagnetic nature of (Ga,Cr)As, the magnetic field dependences of the MCD signals were measured at different temperatures $T$ = 1.5, 5, 8 and 14 K. The temperature dependences of both the spontaneous magnetization $M_S$ and $T_C$ can be determined from the magnetic field dependences of the magnetization characteristics by means of an Arrott plot. Since the Arrott plot analysis only uses data from higher magnetic fields, where the effect of the magnetic anisotropy and the formation of magnetic domains can be neglected, it becomes a reliable method to obtain the values for $M_S$ and $T_C$ [3,14]. In the Arrott plot, the intercept on the $M^2$ axis of a linear extrapolation of the $M^2 \sim B/M$ relation from high magnetic fields corresponds to $M_S^2$. As mentioned previously, the term $\left( \left| M_+ \right|_{vc}^2 - \left| M_- \right|_{vc}^2 \right)$ in Eq. (2) is proportional to the Zeeman spiltting $\Delta E = E_g^- - E_g^+$ near the $\Gamma$ point, and the latter increases as the magnetization $M$ increases. That leads to $MCD^2 \propto M^2$ as well. As a result, Fig. 3 (a) uses the magneto-optical Arrott plot, $MCD^2 - B/MCD$, to estimate $T_C$. Intuitively, the intercepted value on the $MCD^2$ axis decreases as the temperature rises from 1.5 K to 5 K and 8 K. At 14 K it becomes zero. The same procedure is used for the $M^2 \sim B/M$ plot obtained from SQUID for the four different conditions: 1) 5K, $B // [001]$ (closed squares), 2) 5K, $B // [110]$ (closed circles), 3) 10K, $B // [110]$ (closed triangles), 4) 30K, $B // [110]$ (closed stars). Fig. 2 (c) plots the intercept values of $MCD|_{B=0}$ and $M$s from both Fig.2 (a) and (b) against the temperature, revealing a consistent trend. Following a well accepted argument, the curve of $M_S$-$T$ should obey the Brillouin function only if (Ga,Cr)As is a real ferromagnetic semiconductor [3,14]. The solid curve is fitted to the data when $S = 2$ ($Cr^{2+}$), $g = 2$ are assumed [10]. The excellent fitting is not only in accordance with what is expected from a ferromagnetic semiconductor, but also gives an accurate estimate for the Curie temperature, which is $T_C = 12$ K in our (Ga,Cr)As sample with $x = 1\%$.

Thus far, we have not obtained the dependence of the Curie temperature on the mole fraction $x$ of Cr for the low-temperature grown (Ga,Cr)As layers because of the poor crystalline quality of (Ga,Cr)As layers with $x > 1\%$, as already seen in Fig. 1 (a). The remarkable feature that peaked at $\lambda = 890$ nm in the MCD spectrum, as seen in the lower-left inset of Fig. 2, still remains to be understood. One can speculate that this spectral feature is related to the details of the band structure of (Ga,Cr)As, but, this is difficult to be known from the experimental viewpoint up to now.

For the moment, the magnetic and transport properties of (Ga,Cr)As are still far worse than (Ga,Mn)As [9-11]. Nevertheless, the present study on the magneto-optical effect of the (Ga,Cr)As layers provides unambiguous evidence for its intrinsic ferromagnetism with all the necessary properties in accordance with that of a DMS. The prospect of (Ga,Cr)As such as the potential high $T_C$, III-V DMSs might be brilliant from both the theoretical predictions and experiments [8]. Moreover, M. Yamada *et al.* already reported a streaky RHEED pattern in the growth of (Ga,Cr)As with $x = 19\%$ [11]. Meanwhile, in contrast, a $T_C$ over 250 K in (Ga,Mn)As with Mn% = 8% is unlikely under the renewed theoretical prediction [8]. Like (Ga,Mn)As, more effort should be devoted to seek the appropriate conditions for growing (Ga,Cr)As layers before it becomes an appropriate candidate to complement or substitute for (Ga,Mn)As.

In summary, a systematic study of the magnetic circular dichroism (MCD) has been carried out for a series of (Ga,Cr)As epilayers, grown by the low-temperature molecular-beam epitaxy technique. The measured MCD spectra, the magnetic field dependence of MCD and its hysteresis, and the temperature dependence of the spontaneous magnetization all coincide with the typical behaviors of the standard DMS-like (Ga,Mn)As. That convinces us that the (Ga,Cr)As epilayers are intrinsic ferromagnetic semiconductor.

The authors acknowledge J. J. Deng, W. Z. Wang, Y. T. Wang, Z. C. Niu and F. H. Yang for their help with the experiments. This work was in part supported by National Basic Research Program of China No 2006CB932801and No2007CB924904, and also by Special Research Programs of Chinese Academy of Sciences and the National Natural Science Foundation of China under Grant No. 10334030, 10425419, and 60521001.

References

1. A. H. Macdonald, P. Schiffer and N. Samarth, nature materials **4**, 195 (2005).

2. K. Ando, Science **312**, 1883 (2006).

3. K. Ando, H. Munekata, J. Magn. Magn. Mater. **272**, (2004).

4. K. Ando, T. Hayashi, M. Tanaka, A. Twardowski, J. Appl. Phys. **83**, 6548 (1998).

5. H. Saito, V. Zayets, S. Yamagata, K. Ando, Phys. Rev. Lett. **90**, 207202 (2003).

6. K. Ando, Appl. Phys. Lett. **82**, 100 (2003).

7. J. R. Neal, A. J. Behan, R. M. Ibrahim, H. J. Blythe, M. Ziese, A. M. Fox, and G. A. Gehring, Phys. Rev. Lett. **96**, 197208 (2006).

8. J. L. Xu, M. van Schilfgaarde, and G. D. Samolyuk, Phys. Rev. Lett. **94**, 097201 (2005); K. Sato, P. H. Dederics, and H. Katayama-Yoshida, Europhys. Lett., **61**, 403 (2003); L. M. Sandratskii and P. Bruno, Phys. Rev. B **67**, 214402 (2003); Y. Z. Zeng and M. C. Huang, Chin. Phys. Lett. **21**, 1632 (2004).

9. H. Saito, W. Zaets, R. Akimoto, K. Ando, Y. Mishima, and M. Tanaka, J. Appl. Phys. **89**, 7392 (2001).

10. A. Dakhama, B. Lakshmi, and D. Heiman, Phys. Rev. B **67**, 115204 (2003); A. Dakhama, C. Feinstein, and D. Heiman, J. Supercond. **16**, 59 (2003).

11. M. Yamada, K. Ono, M. Mizuguchi, J. Okabayashi, M. Oshima, M. Yuri, H. J. Lin, H. H. Hsieh, C. T. Chen, and H. Akinaga, J. Appl. Phys. **91**, 7908 (2002); K. Kanai, J. Okabayashi, S. Toyoda, M. Oshima, and K. Ono, Appl. Phys. Lett. **88**, 192506 (2006); H. Ofuchi, M. Yamada, J. Okabayashi, M. Mizuguchi, K. Ono, Y. Takeda, M. Oshima, H. Akinaga, Physica B **376-377**, 651 (2006); H. D. Gan, H. Z. Zheng, J. F. Bi, Y. Ji, B. Q. Sun, G. R. Li, K. Zhu, J. H. Zhao, The Fourth International School and Conference on Spintronics and Quantum Information Technology, June 17-22, 2007, Hawaii, USA., p. 50.

12. M. Seefeldt, A. Heuer, and R. Menzel, Opt. Commun. **216**, 199 (2003).

13. See, for example, "Magneto-Optics", edt. by S. Sugano and N. Kojima; " Faraday and Kerr Effects in Ferromagnets" K. Shinakawa ,pp137; Springer-Verlag 1999.

14. F. Matsukura, H. Ohno, A. Shen, and Y. Sugawara, Phys. Rev. B **57**, R2037 (1998).

Figure captions

Fig. 1(a) Double crystal x-ray diffraction measurement under the (004) Bragg reflection for (Ga,Cr)As epilayers of different mole fractions of Cr. (b) Temperature dependence of in-plane remanent magnetization for (Ga,Cr)As samples with $x = 1\%$, measured by SQUID. The inset shows the magnified hysteresis loop near zero magnetic field.

Fig. 2 The magnetic field dependence of normalized MCD at $\lambda = 800$, 890 nm. The lower-left inset is MCD spectra at $B = 1$ T and $T = 1.5$ K for the (Ga,Cr)As sample with $x = 1\%$. The upper-right inset is a comparison between SQUID measured magnetization and MCD characteristics plotted against the magnetic field at $T = 5$ K.

Fig. 3 (a) Arrott plots of MCD intensity at $\lambda = 860$ nm and different temperatures for the (Ga,Cr) As sample with $x$ = 1%. (b) Arrott plots of magnetization measured by SQUID. (c) $MCD|_{B=0}$ and $M$s from both of Fig.2 (a) and (b) plotted as a function of the temperature. The solid curve is a Brillouin function fitting to the data when $S$ = 2 ($Cr^{2+}$), $g$ = 2 are assumed.

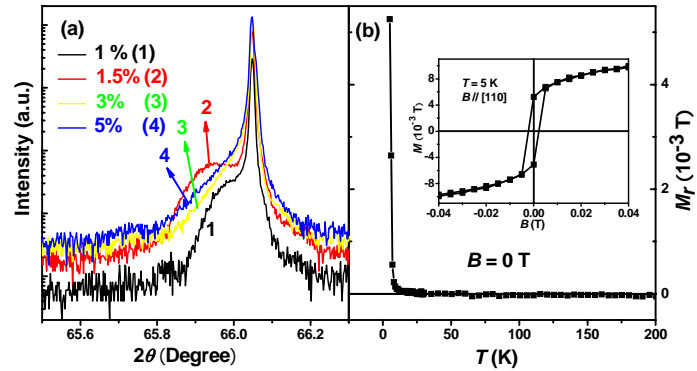

Fig. 1

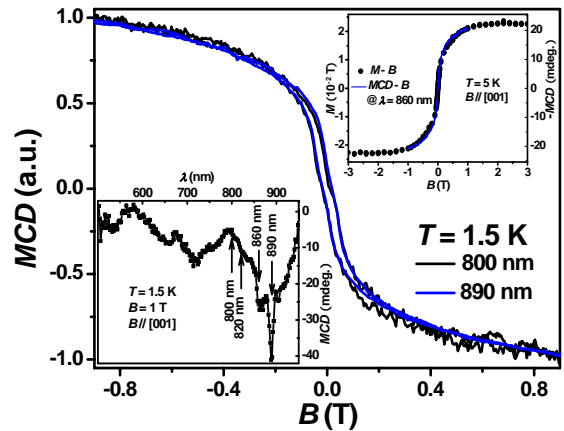

Fig. 2

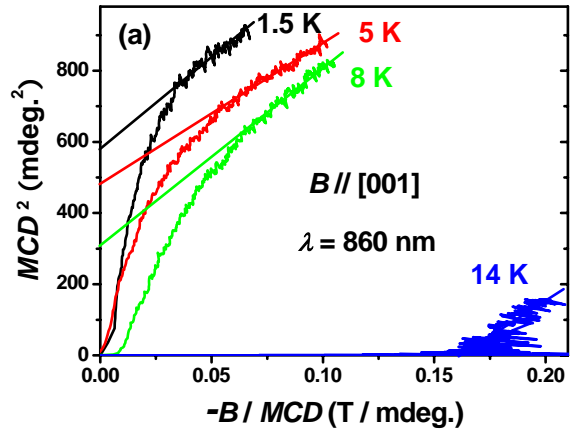

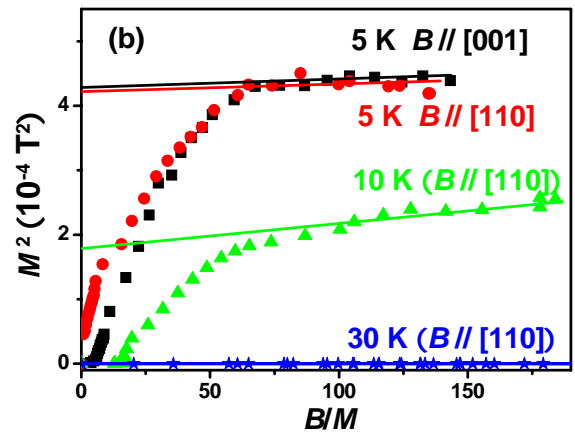

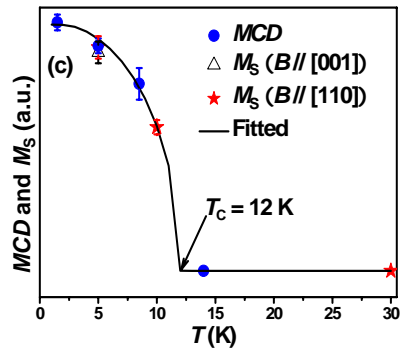

Fig. 3